\documentclass[longbibliography,pra,letterpaper,twocolumn,showpacs,amsmath,amssymb,floatfix]{revtex4-1} 

\usepackage[table]{xcolor} 
\usepackage[utf8]{inputenc} 
\usepackage{graphicx}
\usepackage{bm}
\usepackage{color}

\newcommand{\rmd}{\mathrm{d}}
\newcommand{\rme}{\mathrm{e}}
\newcommand{\rmi}{\mathrm{i}}

\providecommand{\openone}{\leavevmode\hbox{\small1\kern-3.8pt\normalsize1}}

\newcommand{\mcH}{\mathcal{H}}

\newcommand{\fref}[1]{fig.~\ref{#1}}  
\newcommand{\Eref}[1]{Eq.~(\ref{#1})} 
\newcommand{\Sref}[1]{Sec.~\ref{#1}}
\newcommand{\Fref}[1]{Fig.~\ref{#1}}

\newcommand{\env}{\text{e}}	
\newcommand{\h}{\text{H}}	
\newcommand{\sys}{\text{s}}

\newcommand{\eps}{\varepsilon}	
\newcommand{\tr}{\mathop{\mathrm{tr}}\nolimits}
\newcommand{\diag}{\mathop{\mathrm{diag}}\nolimits}
\newcommand{\<}{\langle}
\renewcommand{\>}{\rangle}
\def\1{{\mathchoice{\rm 1\mskip-4mu l}{\rm 1\mskip-4mu l}{\rm 1\mskip-4.5mu l}{\rm 1\mskip-5mu l} }}

\newcommand{\Hs}{\mathcal{H}_\sys}

\newcommand{\He}{H_\rme}

\newcommand{\One}{\openone}

\newcommand{\la}{\langle}
\newcommand{\ra}{\rangle}

\graphicspath{{eps/},{images/},{figures/}}

\begin{document}
\title{Transition from non-Markovian to Markovian dynamics for generic environments} 

\author{Nephtalí Garrido$^1$}
\author{Thomas Gorin$^2$}
\author{Carlos Pineda$^3$}
\address{$^1$ Midlands Ultracold Atom Research Centre, School of Physics and Astronomy, The University of Nottingham,  University Park, Nottingham NG7 2RD, United Kingdom}
\address{$^2$ Departamento de F\'\i sica, Universidad de Guadalajara,
 Blvd. Marcelino Garc\'\i a Barragan y Calzada Ol\'\i mpica, Guadalajara
 C.P. 44840, Jalisco, M\' exico.}
\address{$^3$ Instituto de Física, Universidad Nacional Autónoma de México, México D.F. 01000, México}

\begin{abstract}
Using random matrices, we study the reduced dynamics of a two-level system
interacting with a generic environment. In the weak-coupling limit, the result
can be obtained directly from known results for purity
decay, and result in Markovian dynamics.
We then focus on the case of strong coupling, when the dynamics is known
to be non-Markovian. In this regime,
the coupling dominates over the local parts of the Hamiltonian, and
thus we treat the latter as a perturbation of the former. With the help of a
linear response approximation, this allows us to obtain an analytical
description of the reduced dynamics. Finally, we find a transition from
non-Markovian to Markovian dynamics at a point where the coupling and the
local Hamiltonian are comparable in size.
\end{abstract}
\maketitle
\section{Introduction} 

In Ref.~\cite{PhysRevLett.107.0804042} it was shown that one should
expect non-Markovian behavior when a central system is coupled strongly to a
generic environment. 
In that work everything else but the coupling operator was
neglected~\cite*{GS02b,GS03}.
In the present paper, we will study the fate of non-Markovianity, when the
coupling to the environment is still strong, but a local part is also
present. 
The main mathematical tool to address these questions is random matrix theory
(RMT). This theory has found a wide variety of applications in several
fields~\cite{GMW98}, including quantum chaos, where a direct link between
the ensembles studied in RMT and classically chaotic systems has been well
established~\cite{BerTab77,CVG80,BGS84,BluSmi88}. Moreover, the idea of 
complicate interactions, has been extrapolated to encompass interactions 
between two systems, an idea which was formalized, under certain conditions, by 
Lutz and Weidenm\" uller~\cite{LutWei99}. This can be exploited to, say, develop a 
theory of decoherence with RMT; see~\cite{2008NJPh...10k5016G,PSZ03}.
Considering the coupling term as the unperturbed system,
and the local (free) Hamiltonian as the perturbation, we find a critical
perturbation strength, beyond which the system becomes Markovian. At this point
the free part is equally important as the coupling part.

While in the infinitely strong coupling case (i.e. without local
terms)~\cite{PhysRevLett.107.0804042} it
was possible to obtain an exact analytical solution, here we have to resort to
a linear response approximation~\cite{thomas_fidelity2006}. Even then, the
analytical solution is quite involved, as it requires the calculation of a 
large number of monomial integrals over the unitary group (for simplicity, we 
will assume the absence of any symmetries, including anti-unitary 
ones)~\cite{mello80,gorin2008}. 

The paper is organized as follows:
In the following section, we will describe the system and environment, and show
that the dynamics of the central two-level system is completely determined by a
single real function $\alpha(t)$. We describe the measure of non-Markovianity
which we are using, and review known results of the system in the limit of
strong~\cite{PhysRevLett.107.0804042} and weak 
coupling~\cite{PGS07,2008NJPh...10k5016G}.
Next, in \Sref{sec:small:coupling}, we use the results for the evolution of
purity to calculate the channel for weak coupling.  In \Sref{sec:strong}, we
calculate the linear response approximation for
$\alpha(t)$, 
when both the free part and the coupling term are present.
We obtain an explicit expression when the
dimension of the environment is finite,
and a much simpler one in the infinite case.
We then compare our results to numerical simulations. In \Sref{sec:noma}, we 
discuss the sharp transition between non-Markovian and Markovian dynamics 
halfway between strong and weak coupling, in the limit of 
infinite dimension, where the dimension of 
the environment and the corresponding Heisenberg time are both going to 
infinity.
We finish the paper with \Sref{sec:concl}, in which the conclusions are 
given.

\section{The system} 
Consider the usual system-environment setting, 
with the Hilbert space 
being factored in
\begin{equation}
\mcH = \Hs \otimes \He,
\label{eq:hilbert:space}
\end{equation}
where $\Hs$ corresponds to the system and $\mcH_\env$ to the environment.
Moreover, let us choose a single two-level system (qubit) as central system, 
such that $\dim \mcH_\sys = 2$, and a finite dimensional environment with
$\dim \mcH_\env = N$. 
The Hamiltonian governing the system is set to be
\begin{equation} 
H=s \openone_2\otimes H_\env + V .
\label{eq:hamiltonian} 
\end{equation} 
This represents the simplest nontrivial choice for the local
part of the Hamiltonian, where any dynamics in the qubit is neglected.
We shall distinguish three regimes: the
fully coupled system, when $s=0$; a strongly coupled regime when 
the norm of the coupling $V$ is comparable to the norm of the 
free Hamiltonian $s H_\env$; and a weak coupling regime when the norm of the
coupling is much smaller than that of the free Hamiltonian.  The evolution of
the qubit is given by 
\begin{equation}
\rho_\sys^{(t)} = \tr_\env \left[ U^t \rho_\sys^{(0)} \otimes \rho_\env U^{-t} 
	\right],
\label{eq:rhoevolution}
\end{equation}
where the evolution operator is $ U^t = \exp(-\imath H t)$. We use the Pauli
basis, to represent the quantum channel induced by \Eref{eq:rhoevolution}. The 
corresponding matrix elements are given by 
\begin{equation}
\tilde \Lambda^{(t)}_{j,k} = \frac{1}{2} \tr\left[\sigma^j\otimes\openone_\env
   U^{t} \sigma^k \otimes \rho_\env U^{-t}\right], 
\label{eq:Lambda}
\end{equation}
where $\sigma^0 = \openone$ and 
$\sigma^{1, 2, 3} = \sigma_{x, y, z}$.
Notice that choosing 
$H_\env$ and $V$ in \Eref{eq:hamiltonian} from unitarily invariant 
ensembles, results in an  ensemble of Hamiltonians $H$ that is invariant under local 
unitary transformations. In the case of the central system, this implies
that after averaging, the channel must be isotropic, so its structure is
\begin{equation}
\Lambda^{(t)}=\< \tilde \Lambda^{(t)}\> =\begin{pmatrix}
1 & 0      & 0      & 0 \\
0 & \alpha(t) & 0      & 0\\
0 & 0      & \alpha(t) & 0 \\
0 & 0      & 0      & \alpha(t)
\end{pmatrix}.
\label{eq:paudepha}
\end{equation}
Here, we introduced the notation $\< \cdot \>$ for averages over the 
ensemble of random matrices. 
In the case of the environment, the above invariance property implies that 
$\Lambda^{(t)}$ does not depend on the initial state $\rho_\rme$ of the 
environment. This allows us to replace $\rho_\rme$ with the maximally mixed 
state and write
\begin{equation}
\alpha(t) = 
\frac{1}{N}
  \< \tr[\sigma^3\otimes \openone_\env U^{-t} \sigma^3\otimes\openone_\env 
   U^t]\>.
\end{equation}
\subsection{\label{ssec:fullcoup} Full coupling} 
The solution to the fully coupled case, corresponding to $s=0$ in 
\Eref{eq:hamiltonian}, has been worked out in detail in
Ref.~\cite{PhysRevLett.107.0804042}. Here, we only recall the most important 
results as they are to be generalized in the present work. 
This allows us to introduce some notations. 
For $s=0$, the quantity to be calculated is 
\begin{equation}
\label{cap:modelo:eq:alpha0}
\alpha_0(t)= \frac{1}{N}
   \left \langle{\rm tr}\Big [\, \sigma_z\otimes\One\; 
      \rme^{-\imath V\, t}\;
   \sigma_z\otimes\One\; \rme^{\imath V\, t}\, \Big ]\right \rangle .
\end{equation}
Recall that $V$  is just the 
coupling term, to be chosen from the Gaussian unitary ensemble (GUE) of 
dimension $2N$. We shall diagonalize $V$ (and thus  the evolution operator)
with the unitary
matrix $O$. We thus have
\begin{equation}
U^t = \rme^{-\imath V\, t} = O \mathrm{diag} (\rme^{-\imath v_j t}) O^\dagger ,
\label{eq:splitU}
\end{equation}
where the $\{ v_j\}_j$ are the eigenvalues of $V$. 
We use units for time and energy such that $\hbar$ is eliminated and
the spectral range of $V$ is equal to (unless stated otherwise, the
level density for $V$ obeys a semicircle law). As a result, energies
and times are denoted by dimensionless quantities.
One then 
averages with respect to $O$, with the Haar measure, as explained
in~\cite{mello80,Collins}, and obtains the general expression
\begin{equation}
\alpha_0(t)= \frac{4 N^2 |f(t)|^2 - 1}{4 N^2 - 1},
\label{cap:modelo:eq:alfa_exacta}
\end{equation}
where $f(t) = \frac{1}{2N} \sum_j \exp (-\imath v_j t)$ is the Fourier
transform of the spectral density of the Hamiltonian (remember
that, for $s=0$, $V$ is the Hamiltonian of the system). Notice that this
expression is valid for any unitarily invariant ensemble, not just the GUE. 
One can rewrite the above expression as 
\begin{equation}
\label{cap:modelo:eq:alphague}
\alpha_0(t) =
   \frac{4 N^2 b_1^2(t) + 2N[1-b_2(t)]-1}{4 N^2 -1}\,,
\end{equation}
where $b_1$ is the Fourier transform of the level density of $V$, 
and $b_2$ is the two-point form factor without unfolding; cf. 
Ref.~\cite{mehta}.
For the GUE, both functions are known analytically and are given
in Appendix~\ref{cap:details:s:0} in Eqs.~(\ref{eq:def:b1}) 
and~(\ref{eq:def:b2}) (together with further details).

Spectral correlations are expected to be limited to an energy scale of the 
order of the mean level spacing, which is $N$ times smaller than the energy 
scale of the level density. As a consequence, the relevant time scales for 
$b_1$ and $b_2$ become very different for large dimensions. We chose matrices 
$V$ from the GUE such that $\la V_{ij} V_{kl}\ra = \delta_{jk}\delta_{il}/N$. 
In this way, in the limit $N\to\infty$, the level density tends to a 
semi-circle on the interval $(-1,1)$. As we have set $\hbar = 1$, the relevant 
timescale for $b_1$ is therefore of order 1 (we call this timescale 
``macroscopic''), while for $b_2$ the relevant timescale is the Heisenberg time 
which is of order $N$.
In the limit $N\to\infty$, we get for the GUE an oscillating function in time:
\begin{equation}
\label{cap:modelo:eq:alphague:infinito}
\lim\limits_{N \rightarrow \infty}  \alpha_0(t)
  = \left [\frac{J_1(2\,t)}{t}\right ]^2.
\end{equation}

\begin{figure} 
\centering
\includegraphics{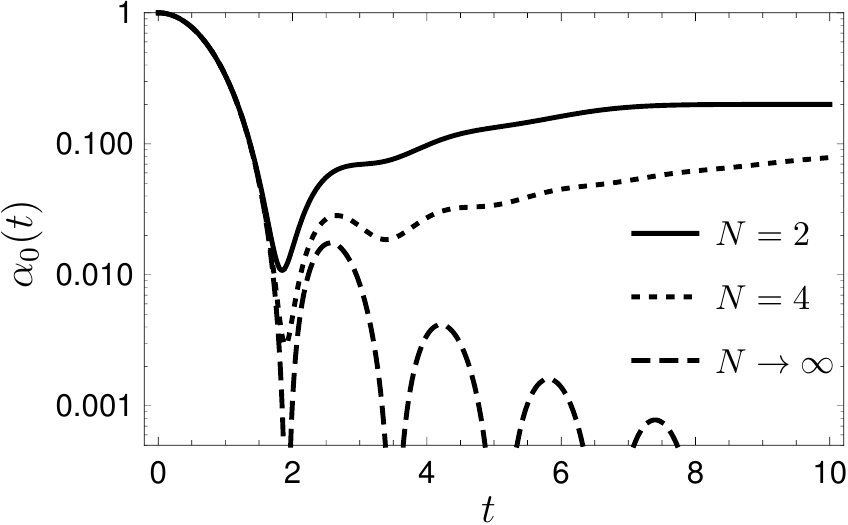}
\caption{$\alpha_0(t)$ as a function of time 
for different dimensions $N$. Nonmonotonic behavior, causing non-Markovianity,
is observed even in the limit of large dimensions, $N\to\infty$.
Here, as well as in all subsequent figures, time is measured in
dimensionless units, as explained in the main text below Eq.~(\ref{eq:splitU}).
}
\label{cap:modelo:fig:Solucion_s=0}
\end{figure}  

\subsection{Non-Markovianity in the fully coupled system} 

Quantum non-Markovianity does not have a unique definition. 
Definitions include considering any deviation from the Lindblad equation as 
non-Markovian behavior~\cite{PhysRevLett.101.150402}, the backflow of
information from the environment into the system~\cite{PhysRevA.81.062115}, and
also the impossibility of defining an instantaneous 
quantum map for intermediate times~\cite{Rivas2010}. Accordingly, several 
measures have been proposed to quantify the degree of non-Markovianity,
each with different properties and problems~\cite{reviewbreuer, reviewrivas}. 
However, for simple channels, like a depolarizing channel, 
as in our case, most definitions coincide as far as the 
distinction between Markovianity and non-Markovianity is
concerned~\cite{PhysRevLett.107.0804042}, even though the measures of the
degree of non-Markovianity 
are usually not comparable.  
For the sake of definitiveness we shall use the measure proposed
in~\cite{PhysRevA.81.062115}, although other measures could be easily 
incorporated in this framework. 
The measure is defined as the maximum of the integrated backflow of 
information measured in terms of increasing distinguishability, where the 
maximum is taken over all possible pairs of initial states. In the present 
case, where the quantum process is determined in terms of the function 
$\alpha(t)$, one gets~\cite{PhysRevLett.107.0804042}
\begin{equation}
\mathcal{M}= 2\int_{\dot\alpha > 0}^{\infty} \rmd t\; 
   \dot\alpha(t) \; .
\label{NMmeasure}\end{equation}
The measure will be greater than zero if and only if 
$\dot \alpha(t) > 0$ for some time, i.e. if the Bloch sphere expands
during a time interval. 
One of the results of Ref.~\cite{PhysRevLett.107.0804042} says that the 
system will {\it generically} display non-Markovianity, even in the limit of
an infinite dimensional environment ($N\to\infty$);
see~\Fref{cap:modelo:fig:Solucion_s=0}. One may compare the present model to
the case of an environment modeled by a collection of harmonic oscillators,
characterized by a spectral density $J(w)$; see~\cite{BrePet02}, chap.~10.
In those models this spectral density has a role
  similar to the level density in ours
(however, see~\cite{prl:88:170407, pra:75:052108}),
as it is the forms of those functions which determine the
reduced dynamics and 
thereby the (non-)Markovianity. This similarity is surprising, 
since we are dealing with a very strong coupling limit, whereas the 
description based on the spectral density  relies on a weak 
coupling approximation. 
In this respect, we also find it surprising that
at 
strong but finite coupling, our model shows a transition to Markovian dynamics, 
independent of the level density. That case will be discussed in 
Sec.~\ref{sec:strong}.

%

At finite $N$, 
the non-Markovianity has two contributions acting at different time scales. The 
first comes from the oscillations in the one-point function $b_1(t)$, which 
appear on a timescale independent of the dimension $N$ of the 
environment (the macroscopic timescale). The second contribution comes from a
recovery of $\alpha(t)$ between the first zero of $J_1(2t)$ and the long-time 
limit 
\begin{equation}
\lim_{t\to\infty} \alpha(t) = \frac{1}{2N+1}.
\end{equation}
That occurs on the timescale of 
the Heisenberg time $\tau_\h$ of the environment, which is proportional to $N$. 
In the semiclassical limit, $N\to\infty$, the Heisenberg time goes to infinity 
and the recovery goes to zero. As a consequence, 
$\lim_{N\to\infty} \mathcal{M} = 0$, also.


\section{The weak-coupling limit} 
\label{sec:small:coupling}
The behavior of $\alpha(t)$ in the weak-coupling limit can be deduced from
previous results~\cite{PGS07,2008NJPh...10k5016G}, where the purity for a model
equivalent to \Eref{eq:hamiltonian} was studied. In that limit, the relevant
time scale is the Heisenberg time $\tau_\h$ of the Hamiltonian $H_\env$ of the
environment. Since the focus was then put on the evolution of purity,
$P=\tr\rho^2$, we use the fact that purity can be expressed in terms of
$\alpha(t)$ from \Eref{eq:paudepha} as follows:
\begin{equation}
P(t)=\tr \left( \Lambda^{(t)}[\rho] \right)^2 = \frac{1+\alpha(t)^2}{2}. 
\label{eq:purityalpha}
\end{equation}
Switching from the parameter $s$, which scales the factorized term, to $\lambda$
scaling the coupling, we can write
\begin{equation}
H=H_\env\otimes\openone + \lambda V .
\label{eq:hamiltonianweak}
\end{equation}
$V$ is chosen from a GUE, but now with an $N$-independent scaling 
$\la V_{ij} V_{kl}\ra = \delta_{jk}\delta_{il}$. In order to map this 
Hamiltonian on \Eref{eq:hamiltonian}, we would have to set 
$\lambda = 1/(sN)$.
In the case that $H_\env$ and $V$ are both members of a GUE, it was found that, 
in the linear response approximation, the average purity is given by
\begin{equation}
P_\text{LR}(t) = 1 - \lambda^2 g(t)
\label{eq:puritylinearresponse}
\end{equation}
with 
\begin{equation}
g(t) = 2t \max\{t, \tau_\h\} +\frac{2}{3\tau_\h}(\min \{t, \tau_\h\})^3,
\label{eq:g}
\end{equation}
with $\tau_\h$ being the Heisenberg time of $H_\env$. 

To go beyond the reach of linear response theory, we exponentiate the result,
such that (i) the first two terms in a Taylor series coincide with the linear 
response result and (ii) the asymptotic value coincides
with a theoretical expectation. Such a heuristic procedure, known as 
{\it exponentiation}, has lead to excellent results~\cite{thomas_fidelity2006}. 
In our case, the procedure leads to
\begin{equation}
P_\text{ELR}(t) = \frac12 +\frac12 
   \exp\left[ \frac12 ( P_\text{LR}(t) - 1)\right].
\label{eq:exponentiation}
\end{equation}
Relying on self-averaging, which is often the case in these kind of 
systems~\cite{2008NJPh...10k5016G}, one can reconstruct $\alpha(t)$ for 
moderate values of the perturbation. Thereby, we obtain
\begin{equation}
\alpha(t) = \exp\left( -\frac{\lambda^2}{2} g(t) \right).
\label{}
\end{equation}
Notice that when the Heisenberg time becomes infinite, we obtain an exponential 
decay for $\alpha(t)$, a result also known as the Fermi golden rule. Notice
also that $g(t)$ as defined in Eq.~(\ref{eq:g}) is a monotonically increasing 
function, which implies that $\alpha(t)$ is monotonically decreasing. This means
that the corresponding dynamics is Markovian, independent of the 
shape of the level density.

\section{The strongly coupled system } 
\label{sec:strong}

So far we found that, for $N\to\infty$, the fully coupled system ($s=0$) shows
non-Markovian dynamics, while at weak coupling, the system becomes Markovian. 
In this section, we consider the crossover region, when $s$ is small but 
finite. The linear response theory developed below is applicable as long as 
$s \lesssim 1$ when the free evolution term and the coupling in the Hamiltonian
in \Eref{eq:hamiltonian} are of the same size. From a technical point of view,
the calculation is much more demanding than usual, because the linear response
expansion is around the fully coupled case.

\subsection{Linear response theory} 
\label{sec:calculation:general}
We will calculate 
\begin{equation}
\alpha(t) = \frac{1}{N} \left\langle
\tr\left[\sigma_z\otimes\openone_\env \rme^{-\imath H t} 
   \sigma_z\otimes\openone_\env \rme^{\imath H t} \right]
\right\rangle
\end{equation}
with the ensemble defined in \Eref{eq:hamiltonian}.
To apply linear response theory for small $s$, we consider the unperturbed
propagator to be $\rme^{-\imath V t}$, and the perturbation $s H_\rme$.
Hence, we have for the echo operator:
\begin{multline}
\rme^{\imath V t}\; \rme^{-\imath H t} \approx \One - \imath s\int_0^t\rmd t'
       \tilde H_\rme(t') \\- s^2\int_0^t\rmd t'\int_0^{t'}\rmd t'' 
      \tilde H_\rme(t')\tilde H_\rme(t'')  ,
\label{eq:echo}
\end{multline}
where $\tilde X (t)= \rme^{\imath V t} X \rme^{-\imath V t}$ denotes the
interaction picture of operator $X$. After some calculations, detailed in 
Appendix~\ref{cap:details:linear}, we find that
\begin{equation}
\alpha(t) \approx 
\alpha_0(t) - s^2\alpha_2(t)
\label{eq:alpha:linear:response}
\end{equation}
where
\begin{equation}
\alpha_2(t) =  \frac{2}{N} {\rm Re}
   \int_0^t\rmd t'\int_0^{t'}\rmd t'' 
      \left( A^{(1)} - A^{(2)} \right)
\label{eq:alpha:two}
\end{equation}
and 
\begin{align}
A^{(1)} &= \left\langle {\rm tr}\big [ \rme^{\imath V t} \sigma_z 
         \rme^{-\imath V (t-t')} H_\rme \rme^{-\imath V (t'-t'')}
         H_\rme \rme^{-\imath V t''} \sigma_z \big ]\right\rangle, 
\label{A1express}\\
A^{(2)} &= \left\langle {\rm tr}\big [ \rme^{\imath V (t-t'')} \sigma_z
         \rme^{-\imath V (t-t')} H_\rme \rme^{-\imath V t'} \sigma_z
         \rme^{\imath V\, t''} H_\rme \big ]\right\rangle .
\label{A2express}
\end{align}
\subsection{Averaging over the unitary group} 

We shall work in the eigenbasis of the environmental Hamiltonian, so 
that $H_\rme = \diag \varepsilon_k$. Let us call $O$ the matrix of
eigenvectors of $V$ so that $\rme^{\imath V t} = O\rme^{\imath\vec{v} t
}O^{\dagger}$, with $\vec{v}$ being the eigenvalues of $V$.  Since $V$ is taken
from a GUE, $O$ must be an element of the unitary group $U(2N)$ equipped with 
the Haar measure. \Eref{A1express} may be rewritten as
\begin{multline}
\la A^{(1)}\ra 
 = \Big\la\rme^{\imath t (v_\alpha- v_\beta) + \imath t' (v_\beta - v_\gamma)
      + \imath t'' (v_\gamma -v_\delta)}\Big\ra \la \eps_k \eps_j\ra
      (-)^{a+d}\\
  \times
    \Big\la O_{dl,\alpha} O_{ai,\alpha}^* O_{ai,\beta} O_{bj,\beta}^*
            O_{bj,\gamma} O_{ck,\gamma}^* O_{ck,\delta} O_{dl,\delta}^*
    \Big\ra\,.
\label{eq:A1}
\end{multline}
In this equation the Einstein summation convention is used. The indices $a$,
$b$, $c$, and $d$ run through the basis states of the qubit; the indices $i$,
$j$, $k$, and $l$ through those of the environment; and the greek indices
through the $2N$ eigenstates of the coupling operator $V$.
Equation~(\ref{eq:A1}) is composed of three independent parts: The first part
contains time and the eigenvalues of the coupling. The second one, contains the
eigenvalues of the environment Hamiltonian, and the third part contains the 
term $(-)^{a+d}$ and the eigenvectors of $V$. 
The term $A^{(2)}$ can be decomposed similarly. Notice that one can go from
\Eref{A1express} to \Eref{A2express} performing the following substitutions:
\begin{equation}
\begin{array}{cccccc}
A^{(1)} &\to& A^{(2)} & \qquad A^{(1)} &\to& A^{(2)}  \\
t &\to& t'' & \qquad\alpha &\to&\delta \\
t' &\to& t\; & \qquad\beta &\to&\alpha \\
t'' &\to& t'\, & \qquad\gamma &\to&\beta \\
 &  & & \qquad\delta &\to&\gamma 
\end{array}\,.
\label{eq:mapping}
\end{equation}
Using these rules, one can write the analogous expression for 
$A^{(2)}$, starting from \Eref{eq:A1}.
As is well known~\cite{mello80}, averages over the unitary
matrices with respect to the Haar measure are invariant under arbitrary
permutations of columns and/or rows.
Hence, the result of those averages only depends on the question of whether these 
indices coincide among each other or not. We may use this invariance property 
to get rid of the factor $(-)^{a+d}$ as follows: Assume $i\ne l$; then the row 
$ai$ is always different from $dl$ and the group average does not depend on
$a$ and $d$, so the summation over $a$ and $d$ can be factored and yields 
$\sum_{a,d} (-)^{a+d} = 0$. Therefore, we may restrict the summation to the 
case $i=l$. 

The different terms in the summation in Eq.~(\ref{eq:A1}) can be grouped 
according to the degeneracy of the indices; the particular value of each index 
is unimportant. One can therefore divide the set of values for the four Greek 
indices into 15 different partitions, which will be enumerated as follows:
\begin{equation} 
\begin{array}{ccl}
 1 &:& \alpha = \beta = \gamma = \delta \\ \hline
 2 &:& \alpha = \beta = \gamma \ne \delta \\
 3 &:& \alpha = \beta = \delta \ne \gamma \\
 4 &:& \alpha = \gamma = \delta \ne \beta \\
 5 &:& \alpha \ne \beta = \gamma = \delta  \\ \hline
 6 &:& \alpha = \beta \ne \gamma = \delta \\
 7 &:& \alpha = \gamma \ne \beta = \delta \\
 8 &:& \alpha = \delta \ne \beta = \gamma 
\end{array} \;  \qquad
\begin{array}{ccl}
 9 &:& \alpha = \beta \ne \gamma \ne \delta \\
 10 &:& \alpha = \gamma \ne \beta \ne \delta \\
 11 &:& \alpha = \delta \ne \beta \ne \gamma \\
 12 &:& \alpha \ne \delta \ne \beta = \gamma \\
 13 &:& \alpha \ne \gamma \ne \beta = \delta  \\ 
 14 &:& \alpha \ne \beta \ne \gamma = \delta \\ \hline
 15 &:& \alpha \ne \beta \ne \gamma \ne \delta
\end{array} \; .
\label{eq:fifteen}
\end{equation} 
For the latin index pairs we can proceed likewise. Due to the invariance 
properties of the averages of the monomials, based on the above labeling of the 
partitions, we can write
\begin{align}
A^{(1)} &= \sum_{I = 1}^{15}\sum_{J=1}^{15} 
   C_I M_{IJ}^{(1)} F_J^{(1)}
    = \bm{C}^T \bm{M}^{(1)} \bm{F}^{(1)} , \nonumber \\
A^{(2)} &= \bm{C}^T \bm{M}^{(2)} \bm{F}^{(2)}.
\label{A1A2results}
\end{align}
Notice that we are using capital latin letters as indices for the different 
partitions. 
In this equation, $\bm{C}$ is a vector containing all $C_I$ 
cases, in which the terms $\la \eps_j\eps_k\ra$ and $(-)^{a+d}$ are taken into
account; in the matrices $\bm{M}^{(1,2)}$, the group averages over the 
monomials of matrix elements of $O$ are arranged, and the time-dependent phases 
containing the eigenvalues of the coupling are are included in $\bm{F}^{(1,2)}$. 
The partitions, \Eref{eq:fifteen}, with respect to row indices (latin index 
pairs) and column indices (greek indices) have different multiplicities, which 
are included in the vectors $\bm{C}$ and $\bm{F}^{(1,2)}$, respectively.
The factors $C_I$ are the same for $A^{(1)}$ and $A^{(2)}$. 
We find that 
\begin{align*}
C_1 &= -C_2 = -C_5 = -C_6 = -C_7 = 2N,\\
C_3 &= C_4 = -2(N-2), \, C_8 = 2N(2N-1)\\
C_9 &= 4(N-1), C_{10} = -C_{12} = C_{13} = C_{14} = 4(N-1)  \\
C_{11} &= -4(N-1)(N-2), C_{15} = 4(N-1)(N-4).
\end{align*}
The group averages appearing in the matrices $\bm{M}^{(1,2)}$ are
calculated exactly for arbitrary $N$, based on recursion formulas
developed in~\cite{gorin2008}, available as computer code in~\cite{codelibs}. 
We report the results of the vectors $\bm{C}^T \bm{M}^{(1,2)}$:
\begin{widetext}
\begin{align} 
%
\bm{C}^T \bm{M}^{(1)} = \frac{1}{N(2N+1)(2N+3)} \Big(
  N + 4  ,
  \frac{N-1}{2N-1}  ,
  \frac{2(N-1)(N+2)}{2N-1},
\frac{N - 1}{(2 N - 1)}  ,
  \frac{2(N^2+3N+1)}{(2 N - 1)},
 - \frac{N - 1}{N (2 N - 1)}  , \notag\\
  \frac{(N - 1) (N + 2)(2N + 1)}{ N (2N-1)}  ,
 - \frac{N-1}{N (2 N - 1)}  ,
   -\frac{N-1}{2 N (2 N - 1)}  ,
  - \frac{3N+2}{2 N (2 N - 1)},
 - \frac{N - 1}{2N(2 N - 1)}  , \notag\\
-\frac{N-1}{2 N (2 N - 1)},
 \frac{4N^3+6N^2-3N-2}{2N (2 N - 1)}  ,
 -\frac{N-1}{2 N (2 N - 1)}  ,
  \frac{5}{2(2 N - 3) (2 N - 1)} \Big )
\label{eq:ctm1}
\end{align} 
and
\begin{align} 
\bm{C}^T \bm{M}^{(2)} = \frac{1}{N(2N+1)(2N+3)} \Big (
  N + 4  , 
  \frac{N - 1}{2 N - 1}  ,
  \frac{N - 1}{2 N - 1}  ,
  \frac{N - 1}{2 N - 1}  , 
  \frac{N - 1}{2 N - 1}  , 
- \frac{N - 1}{N (2 N - 1)}  ,
  \frac{2 (N - 1) (N + 1)}{N (2N-1)}  ,\notag \\
  \frac{(N+1)(4N +1)}{N (2 N - 1)}  , 
   \frac{2N^2+2N +1}{2 N (2 N - 1)}  , 
\frac{(N-1)(N+1)}{N (2 N - 1)}  , 
   \frac{2N^2 +2N +1}{2N (2 N - 1)}  ,
 \frac{2N^2+2N +1}{2 N (2 N - 1)}  , 
   \frac{(N-1)(N+1)}{N (2 N - 1)}  , \notag\\
\frac{2N^2+2N +1}{2 N (2 N - 1)}  , 
  \frac{2(N-1)(N+1)}{(2 N - 3) (2 N - 1)} \Big ).
\label{eq:ctm2}
\end{align} 
\end{widetext}
\subsection{Average over the eigenvalues of \boldmath $V$} 

We now calculate the components $F^{(1,2)}_I$ of the time-dependent factors
$\bm{F}^{(1,2)}$. As we are mainly interested in the case of large $N$, we 
shall ignore all spectral correlations, as these could only affect the dynamics
of the qubit at times proportional to $N$, where $\alpha(t)$ already is of 
order of $1/N$. We have seen this explicitly in \Sref{ssec:fullcoup}, where we
considered the case of full coupling, $s=0$. We expect that, for the perturbed
case with finite $s$, the situation will be similar, and will be 
justified {\it a posteriori} with the numerical simulations.
In other words, we assume that the eigenvalues of the coupling term $V$ have a
semicircle spectral density, but are otherwise statistically independent. 
Note, however, that in principle, one could take into account correlations and 
describe the behavior up to times of the order of the Heisenberg time, if 
required.

\begin{table} 
\newcommand{\cg}{\cellcolor{green}}
\noindent
\begin{center}
\begin{tabular}{|c|ccc|ccc|}
 \hline
 $\nu$ &$ \left[{\bf C}^T {\bf M}^{(1)}\right]_\nu$& $ F^{(1)}_\nu$  & $\left[A^{(1)}\right]_\nu$&$ \left[{\bf C}^T {\bf M}^{(2)}\right]_\nu$& $ F^{(2)}_\nu$  & $\left[A^{(2)}\right]_\nu$ \\ 
 \hline
 \hline
1     &  -2  &  1  &-1   &-2  & 1  & -1     \\
2     &  -3  &  2  &-1   &-3  & 2  & -1     \\
3     &  -2  &  2  & 0   &-3  & 2  & -1     \\
4     &  -3  &  2  &-1   &-3  & 2  & -1     \\
5     &  -2  &  2  & 0   &-3  & 2  & -1     \\
6     &  -3  &  2  &-1   &-4  & 2  & -2     \\
7     &  -2  &  2  & 0   &-3  & 2  & -1     \\
8     &  -4  &  2  &-2   &-3  & 2  & -1     \\
9     &  -4  &  3  &-1   &-3  & 3  &  0     \\
10    &  -4  &  3  &-1   &-3  & 3  &  0     \\
11    &  -4  &  3  &-1   &-3  & 3  &  0     \\
12    &  -4  &  3  &-1   &-3  & 3  &  0     \\
\cg 13&  -2  &  3  &\cg 1&-3  & 3  &  0     \\
14    &  -4  &  3  &-1   &-3  & 3  &  0     \\
\cg 15&  -5  &  4  &-1   &-3  & 4  & \cg 1  \\
\hline
\end{tabular}
\end{center}
\caption{
\label{cap:pue:tabla:order_A1}
(Color online) Order of magnitude of the 15 different terms [see \Eref{eq:fifteen}] 
contributing to \Eref{eq:alpha:linear:response}. The leading terms are shaded 
(in green). Thus, for large $N$, it is enough to consider the 13th term of 
$A^{(1)}$ and the 15th term of $A^{(2)}$.}
\end{table} 
That said, all components $F^{(1,2)}_I$ will depend solely on the Fourier
transform of the spectral density $b_1$. With the help of the auxiliary 
functions
\begin{align}
\mathcal{F}(x)     & = \frac{(2N)!}{(2N-2)!} b_1^2(x), \nonumber\\
\mathcal{G}(x,y,z) & = \frac{(2N)!}{(2N-3)!}b_1(x) b_1(y) b_1(z), \label{cap:GUE:eq:aux}\\
\mathcal{H}(x,y,z) & = \frac{(2N)!}{(2N-4)!}b_1(x) b_1(y-x) b_1(z-y) b_1(z). \nonumber
\end{align}
we may write
\begin{align} 
\langle F_1^{(1)} \rangle &= 2N,                              &\langle F_9^{(1)}   \rangle&= \mathcal{G}(t',t''-t',t''),   \notag\\
\langle F_2^{(1)} \rangle &= \mathcal{F}(t''),                &\langle F_{10}^{(1)}\rangle&= \mathcal{G}(t+t''-t',t'-t,t''),\notag\\
\langle F_3^{(1)} \rangle &= \mathcal{F}(t'-t''),             &\langle F_{11}^{(1)}\rangle&= \mathcal{G}(t-t'',t'-t,t''-t'),\notag\\
\langle F_4^{(1)} \rangle &= \mathcal{F}(t-t'),               &\langle F_{12}^{(1)}\rangle&= \mathcal{G}(t,t''-t,t''),     \notag\\
\langle F_5^{(1)} \rangle &= \mathcal{F}(t),                  &\langle F_{13}^{(1)}\rangle&= \mathcal{G}(t,-t+t'-t'',t''-t'),\notag\\
\langle F_6^{(1)}   \rangle&= \mathcal{F}(t'),                &\langle F_{14}^{(1)}\rangle&= \mathcal{G}(t,t'-t,t'), \notag\\
\langle F_7^{(1)}   \rangle&= \mathcal{F}(t-t'+t''),          &\langle F_{15}^{(1)}\rangle&= \mathcal{H}(t,t',t'')\,. \notag \\
\langle F_8^{(1)}   \rangle&= \mathcal{F}(t-t''),             &
\label{cap:gue:eq:promf1}
\end{align} 
Finally, using the mapping \eqref{eq:mapping}, we also obtain the components of 
$F^{(2)}$:
\begin{align} 
\langle F_1^{(2)}\rangle&= 2N,                                &\langle F_9^{(2)}\rangle&=\mathcal{G}(t'',t'-t'',t'), \notag \\
\langle F_2^{(2)}\rangle&=\mathcal{F}(t''),                   &\langle F_{10}^{(2)}\rangle&=\mathcal{G}(t'',-t''+t-t',t'-t),  \notag \\      
\langle F_3^{(2)}\rangle&=\mathcal{F}(t'),                    &\langle F_{11}^{(2)}\rangle&=\mathcal{G}(t,t'-t,t'), \notag \\
\langle F_4^{(2)}\rangle&=\mathcal{F}(t-t'),                  &\langle F_{12}^{(2)}\rangle&=\mathcal{G}(t'',t-t'',t), \notag \\
\langle F_5^{(2)}\rangle&=\mathcal{F}(t''-t),                 &\langle F_{13}^{(2)}\rangle&=\mathcal{G}(t''+t'-t,t-t'',t'), \notag \\
\langle F_6^{(2)}\rangle&=\mathcal{F}(t''-t'),                &\langle F_{14}^{(2)}\rangle&=\mathcal{G}(t''-t',t-t'',t'-t), \notag \\
\langle F_7^{(2)}\rangle&=\mathcal{F}(t''-t+t'),              &\langle F_{15}^{(2)}\rangle&=\mathcal{H}(t'',t,t')\,.\notag \\
\langle F_8^{(2)}\rangle&=\mathcal{F}(t),                     &
\label{cap:gue:eq:promf2}
\end{align} 
Equations~(\ref{eq:alpha:linear:response}), (\ref{eq:alpha:two}),
and~(\ref{A1A2results}), together with Eqs.~(\ref{eq:ctm1}) 
to~(\ref{cap:gue:eq:promf2}), provide the final, general result. It is valid,
either in the absence of spectral correlations in $H_\env$, or for large
$N$ at times sufficiently small compared to the Heisenberg time. In our
case, $b_1(t)$ is given in \Eref{cap:modelo:eq:alphague:infinito}, which 
corresponds to a semicircle level density. However, other cases with different 
level densities could be considered, also.
Our general result still requires the evaluation of the double time integral
in \Eref{eq:alpha:two}. Typically, one would have to do this evaluation 
numerically.

\subsection{The solution for large dimensions and times} 

\begin{figure} 
\begin{center}
\includegraphics[width=\linewidth]{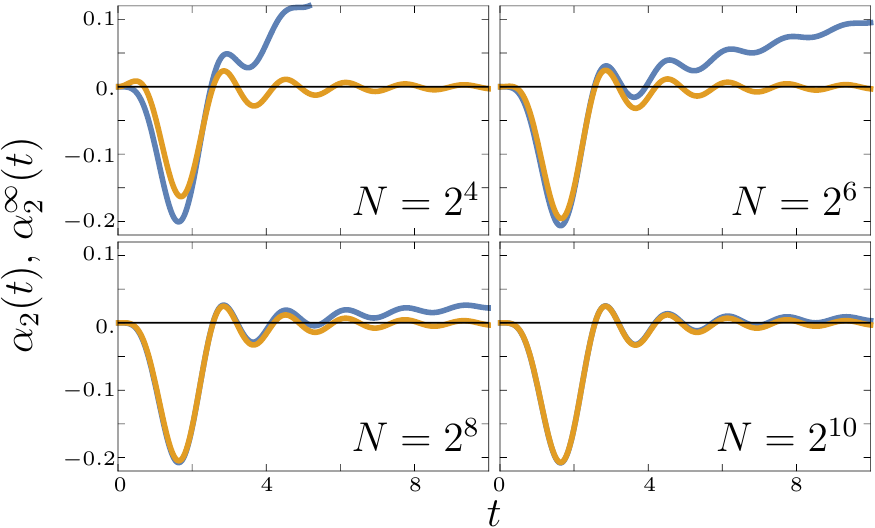}
\end{center}
\caption{\label{fig:alpha_vs_large_n_limit}
(Color online)
Exact value of the integral $\alpha_2(t)$ in \Eref{eq:alpha:two}, with all 
terms (yellow curves) and only taking into account the leading terms in $N$ 
(blue curves); that is, the value of $\alpha_2^\infty(t)$, for several values 
of the dimension $N$. }
\end{figure}  

It is possible to simplify the general expressions, discussed above,
considering explicitly the limit of large $N$. 
Table~\ref{cap:pue:tabla:order_A1} indicates the leading order in $N^{-1}$ of 
all relevant terms, in Eqs.~(\ref{eq:ctm1}), (\ref{eq:ctm2}), and 
(\ref{cap:GUE:eq:aux}). By proper selection of the highest order terms, we
obtain for 
$\alpha_2^\infty(t) = \lim_{N\to \infty} \alpha_2(t)$ the following
\begin{multline}
 \alpha_2^\infty(t) = 2 \int_0^t\rmd t'\int_0^{t'}\rmd t''  
      \bigg( 
     b_1(t)b_1(t'-t-t'')b_1(t''-t') \\
     -
     b_1(t'')b_1(t-t'')b_1(t'-t)b_1(t')
      \bigg).
\label{largeNresult}
\end{multline}
We have tested the reach of this limit in
\fref{fig:alpha_vs_large_n_limit}, where we can see that already for $N=2^{10}$
there is almost no visible difference, for the times reported, between the full
expression and the large-environment limit. Although this 
expression means a considerable simplification for the $b_1$ from a 
semicircle level density, we were still not able to solve the time integrals
in closed form. We found only one case where that is possible. This case, where 
the level density is a Gaussian function, is treated in
Appendix~\ref{cap:PUE:promedios}.

\subsection{The solution for large times} 

Linear response theory is valid whenever the corrections in the echo operator,
\Eref{eq:echo}, with respect to the identity are small; that is, whenever
$|\rme^{\imath V t} \rme^{-\imath H t} - \One | \ll 1$ (here, $|\, \cdot\, |$
denotes the operator norm). This implies that the eigenvalues
of the echo operator must remain close to 1. Departure from that happens for
any value of the perturbation $s$, for sufficiently long times. However, the
smaller the $s$, the larger the time of the validity of the linear response
approximation. 

\begin{figure} 
\includegraphics[width=\columnwidth]{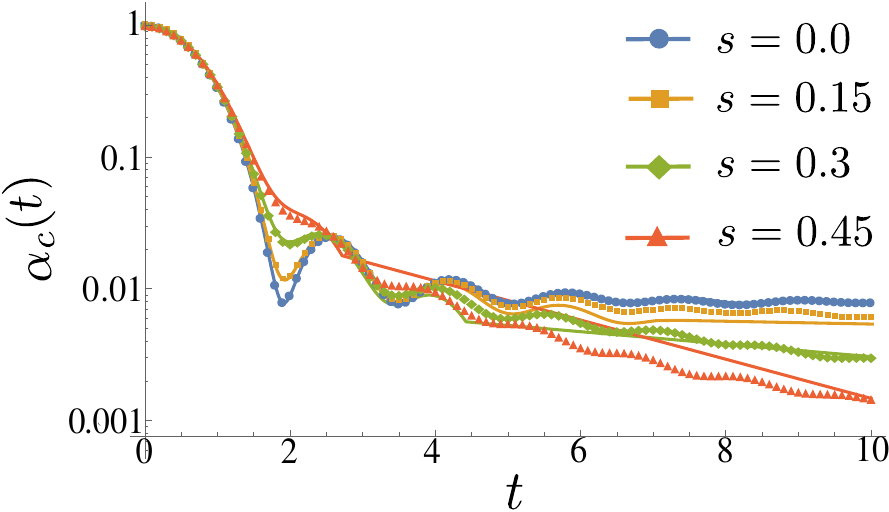}
\caption{
(Color online)
Comparison of the linear response theory, with an exponential tail, and the
numerical simulation, with $N=64$, for several perturbations.}
\label{fig:alphacorte}
\end{figure} 

The extension of linear response formulas in this context is often done using
{\it exponentiation}, as in \Sref{sec:small:coupling}. However, in the present
case such attempts have been unsuccessful~\cite{Garrido:Tesis}. 
As an alternative, we have opted to combine the two linear response theories,
namely the ones discussed in \Sref{sec:small:coupling} and
\Sref{sec:calculation:general}. We shall use the linear response formula
\Eref{eq:alpha:linear:response}
until the time in which the largest (in absolute value) eigenphase reaches
$\pm\pi$.  Afterwards, an exponential decay is fitted. 

\section{\label{sec:noma} The transition from non-Markovian to Markovian behavior} 
\begin{figure}[tb] 
\centering
\includegraphics[width=\columnwidth]{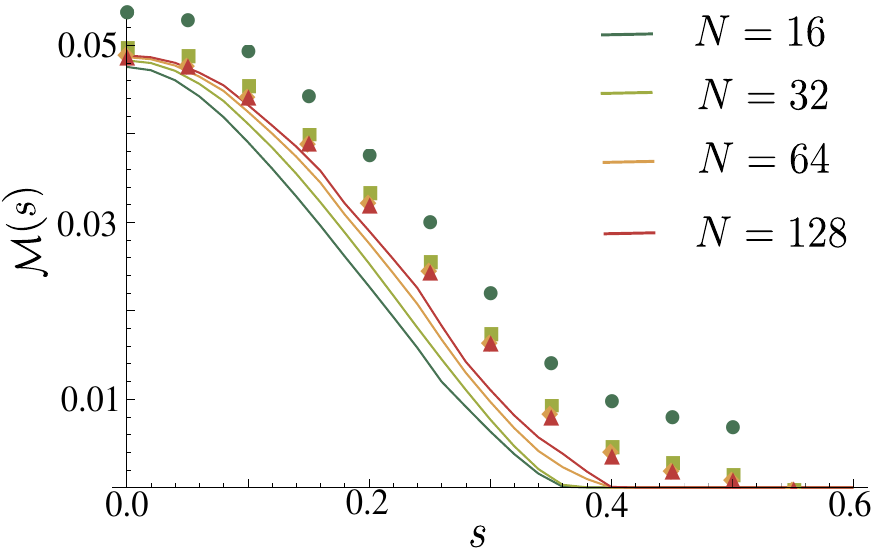}
\caption{
(Color online)
Measure of non-Markovianity for a random Hamiltonian of the 
form \Eref{eq:hamiltonian}. Here, both matrices 
$V$ and $H_\env$ are taken from the GUE, and 
the dimension of the environment $N$ is varied. Symbols indicate calculations
done with the linear response theory, extended with an exponential decay, 
while continuous lines are obtained numerically. One can see that, at a critical
intensity of the coupling $s\approx 0.4$, the system switches from 
non-Markovian to Markovian. It should also be noted that, as the dimension
increases the approximations are more accurate.
\label{fig:no-marko_smc_comparacion}
}
\end{figure}  

As the coupling of the system diminishes (that is, when $s$ increases), one
should fall back to the Markovian case in the large-dimension
limit~\cite{reviewbreuer,2008NJPh...10k5016G}; cf. also 
\Sref{sec:small:coupling}.
This is indeed observed in \fref{fig:alphacorte}, where 
the curves for $\alpha(t)$ become monotonic as $s$ increases.
Thus, we would like to know whether there is a particular value 
for $s$ beyond which the dynamics is Markovian.
This question is answered in \fref{fig:no-marko_smc_comparacion}, where the 
measure for non-Markovianity, 
from Eq.~(\ref{NMmeasure}) is plotted as a function of the coupling $s$.
The points mark the numerical results for $\mathcal{M}$ where 
the integration in Eq.~(\ref{NMmeasure}) has been restricted to the interval
$t\in [0,10]$. While this introduces an error at small dimensions, this error
vanishes at large $N$. The solid lines mark the same quantity, but calculated 
from the composite linear response results shown in \fref{fig:alphacorte}.
One can observe that there is a seemingly sharp transition in the
large-dimension limit, which is not observed for smaller dimensions due to the
two-point correlations that cause an increase in the function $\alpha$, and are
not 
taken into account in the linear response results. 

It is remarkable that a critical value of the coupling is needed to go 
from one regime into the other. It must be noticed, however, that in
our calculation we are using an ensemble of Hamiltonians to 
describe the environment and the coupling to it. In a real experiment this
would correspond to measurements which require many repetitions of the quantum
process, during which the dynamics in the environment changes, e.g., due to 
fluctuating classical fields. 
A particular member of the ensemble will exhibit oscillations that will result 
in non-Markovianity. However, one should distinguish oscillations due to the 
particularities of the system, from generic oscillations due to general 
features of the whole ensemble. 

\begin{figure}
\centering
\includegraphics[width=\columnwidth]{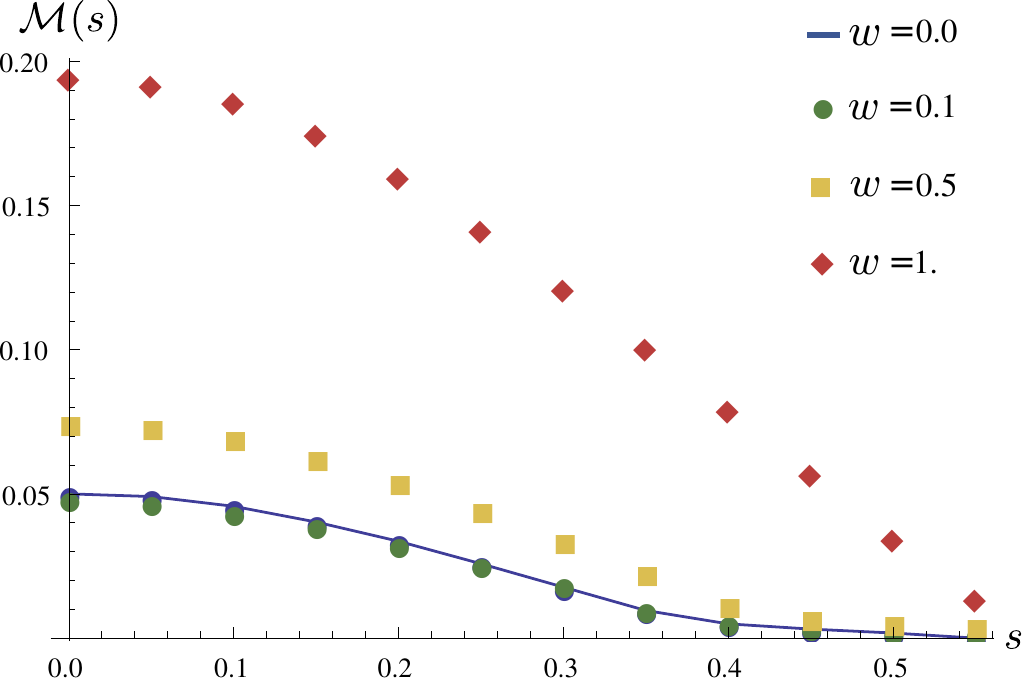}
\caption{The effect of an internal Hamiltonian
in the central system, as in Eq.~\ref{eq:internal}, is studied.
The level splitting, $\omega$, amplifies the non-Markovian effects, but
apparently 
conserves the transition from non-Markovian to Markovian behavior, which occurs
at approximately the same critical value for $s$.
\label{fig:internal}
}
\end{figure}

For completeness, we have also studied the case in which the qubit has an
internal Hamiltonian, where Eq.~(\ref{eq:hamiltonian}) is substituted by
\begin{equation}
H=\omega \sigma_z + s \openone_2\otimes H_\env + V .
\label{eq:internal}
\end{equation}
This Hamiltonian is no longer invariant under unitary transformations in the 
central system, and hence Eq.~(\ref{eq:paudepha}) is no longer valid. Instead,
the new quantum channel will be a combination of a dephasing and a depolarizing
channel.
The only energy scale retained in the limit of large dimensions is the spectral
span of the coupling Hamiltonian $V$ [the level density has the shape of
a semi-circle in the interval $(-1,1)$]. Therefore, one may expect that the 
effect of the additional term depends on the size of $\omega$ as compared to 
the spectral span. Hence, for $\omega \ll 1$ the effect should be negligible,
we do expect differences for $\omega \gtrsim 1$.
In Fig.~\ref{fig:internal}, we present simulations for different values of
$\omega$. The figure shows our measure for non-Markovianity as a function of
$s$, just as in Fig.~4. We can observe, that increasing $\omega$ leads to 
larger values for the measure, but that the transition from non-Markovian to
Markovian behavior is essentially unchanged.

\section{\label{sec:concl} Conclusions} 

We considered a quantum two-level system coupled to a generic 
environment modeled by random matrix theory. We obtained analytical 
expressions for the reduced dynamics using linear response approximations, 
both around the weak and strong coupling limits. 
For the weak-coupling limit, an explicit expression is obtained. 
  The corresponding expression involves a double
   time-integral of one- and two-point functions of the coupling in the 
   general case, a of one-point functions in the large dimension limit.
%
%
In the limit $N\to\infty$, the result becomes much 
simpler [see Eq.~(\ref{largeNresult})]: only two terms survive, which contain 
one-point functions. Nevertheless, the double time-integral still has
to be evaluated numerically.

We then studied the degree of non-Markovianity in the system, using the measure
proposed in~\cite{PhysRevA.81.062115}, based on distinguishability. We show
that the degree of non-Markovianity of the $s=0$ case (infinite coupling) 
considered in Ref.~\cite{PhysRevLett.107.0804042} diminishes as the coupling
term becomes less important, and that in the large-$N$ limit it vanishes at a 
point where the free and coupling terms of the Hamiltonian are of 
equal size.

 
{\em Acknowledgements--} Support by  projects
CONACyT 153190, CONACyT 129309 and  UNAM-PAPIIT IN111015 are acknowledged.
\appendix
\section{Details for the fully coupled case, in the GUE case} 
\label{cap:details:s:0}
The spectral correlations for the GUE are expressed in terms of the functions
$$\phi_j(E) = \frac{\rme^{-2N E^2/4}}{\sqrt{2^j j! \sqrt{2 \pi /(2N)}} }
\mathcal{H}_j (E \sqrt{N}),$$
where  $\mathcal{H}_j$ are Hermite polynomials~\cite{mehta}. 

We fix the normalization so the average of the square of
the diagonal elements in the matrices is $1/N$.
Then, the spectral density for finite dimensions is 
\begin{equation}
R_1(E) = \sum_{j=0}^{2N-1} \phi_j(E)^2
\end{equation}
and the cluster function, containing the correlations between levels, 
is 
\begin{equation}
T_2(E_1, E_2) = \left(\sum_{j=0}^{2N-1} \phi_j(E_1) \phi_j (E_2)\right )^2\,.
\end{equation}
If we define 
\begin{equation}
b_1 (t) = \frac{1}{2N}\int dE \rme^{-\imath E t} R_1(E)
\label{eq:def:b1}
\end{equation}
and
\begin{equation}
b_2 (t) =  \frac{1}{2N}\int dE_1 dE_2 \rme^{-\imath (E_1 - E_2) t} T_2(E_1,E_2)
\label{eq:def:b2}
\end{equation}
we have, for this case, 
\begin{equation}
\langle \alpha_0(t)_{\mathrm{GUE}}\rangle= 
   \frac{4 N^2 b_1^2(t) + 4N(1-b_2(t))-1}{4 N^2 -1}\,,
\end{equation}
since 
\begin{multline}
N^2\<f(t)|^2 \>
=N+\\
\int \rmd E_1 \rmd E_2 \rme^{-\rmi(E_1-E_2)t } [R_1(E_1)R_1(E_2)-T_2(E_1,E_2)].
\end{multline}

In the large dimension limit, we have
\begin{equation}
\label{cap:GUE:eq:charact_sc}
b_1(t) =\frac{J_1(2 t)}{t}\,
\end{equation}
and thus
\begin{equation}
\lim\limits_{N \rightarrow \infty} \langle \alpha_0(t) \rangle_{\mathrm{GUE}}
  = \left [\frac{J_1(2\,t)}{t}\right ]^2.
\end{equation}

\section{Details for the linear response theory} 
\label{cap:details:linear}

Now, we write $\alpha(t)$ in terms of the echo operator as follows:
\begin{equation}
\alpha(t) = {\rm tr}\Big [\, \sigma_z\otimes\One\; \rme^{-\imath V\, t}\; M\;
   \sigma_z\otimes\One\; M^\dagger\; \rme^{\imath V\, t} \, \Big ] \; .
\end{equation}
Using Born approximation \Eref{eq:echo}, we obtain
\begin{align*}
\alpha(t) &= {\rm tr}\big [\, \tilde\sigma_z(t)\; M\; \sigma_z\otimes\One\; 
 M^\dagger\, \big ] \notag\\
 &\approx \alpha_0(t) - s^2
   \int_0^t\rmd t'\int_0^{t'}\rmd t'' {\rm tr} A(t'',t',t),
\end{align*}
where
\begin{equation}
\alpha_0(t)={\rm tr}\left[ \tilde\sigma_z(t) \sigma_z\otimes\One  \right]
\end{equation}
represent the exact known solution for $s=0$, and 
\begin{align*}
A(t'',t',t) &= \tilde\sigma_z(t) \tilde H_\rme(t') \tilde H_\rme(t'')
   \sigma_z - \tilde\sigma_z(t) \tilde H_\rme(t')
   \sigma_z \tilde H_\rme(t'') \notag\\
&\, + \sigma_z \tilde H_\rme(t'') \tilde H_\rme(t') \tilde\sigma_z(t) 
 - \tilde\sigma_z(t) \tilde H_\rme(t'') \sigma_z \tilde H_\rme(t').
\end{align*}
Due to the fact that the matrices $\tilde H_\rme$, $\tilde\sigma_z$ and 
$\sigma_z\otimes\One$ are Hermitian, we obtain the useful identities
\begin{equation*}
{\rm tr}\big [ \tilde\sigma_z(t) \tilde H_\rme(t') \tilde H_\rme(t'')
      \sigma_z \big ]^*
 = {\rm tr}\big [ \sigma_z \tilde H_\rme(t'')
      \tilde H_\rme(t') \tilde\sigma_z(t) ],
\end{equation*}
and
\begin{align*}
{\rm tr}\big [ \tilde\sigma_z(t) \tilde H_\rme(t') \sigma_z \tilde H_\rme(t'')
\big ]^* &= {\rm tr}\big [ \tilde H_\rme(t'') \sigma_z
      \tilde H_\rme(t') \tilde\sigma_z(t)  \big ] \notag\\
 &= {\rm tr}\big [ \tilde\sigma_z(t) \tilde H_\rme(t'')
      \sigma_z \tilde H_\rme(t') \big ]  .
\end{align*}
This implies that the trace of $A(t'',t',t)$ can be written as twice the real
part of the trace of $A_{\rm c}(t'',t',t)$, where the latter quantity only
contains the two terms on the left-hand side of the above equation. We may thus write
for $\alpha(t)$ in the linear response approximation:
\begin{equation*}
\alpha(t) \approx 
\alpha_0(t)
 - 2\, s^2\; {\rm Re}
   \int_0^t\rmd t'\int_0^{t'}\rmd t''\; 
      {\rm tr}\big [\, A_{\rm c}(t'',t',t)\, \big ] \; ,
\end{equation*}
where
\begin{equation*}
A_{\rm c}(t'',t',t) = \tilde\sigma_z(t) \tilde H_\rme(t')
   \tilde H_\rme(t'') \sigma_z
 - \tilde\sigma_z(t) \tilde H_\rme(t') \sigma_z
   \tilde H_\rme(t'')  .
\end{equation*}
Now we split $A_{\rm c}$ in its two parts,
\begin{equation*}
{\rm tr} A_{\rm c}(t'',t',t) = A^{(1)} - A^{(2)},
\end{equation*}
insert identity operators $\rme^{-\imath V\, t}\rme^{\imath V\, t}$ 
wherever necessary, and use the cyclical property of 
the trace, to rewrite more conveniently the term under the
integral.
\section{Normalization of the ensembles considered} 
\label{app:normalization}
In Sec.~\ref{sec:calculation:general}, we are free to consider
any normalization condition. 

We conveniently assume that, with respect to an
arbitrary basis, the matrix elements of both $V$ and $H_\rme$ have a variance
equal to their inverse dimension. Let $H$ be either $V$ or $H_\rme$, so 
the normalization condition reads
\begin{equation}
\la H_{ij}\, H_{kl}\ra = \frac{\delta_{jk}\delta_{li}}{N}\; .
\end{equation}
That implies for the average of the trace of $H^2$ and for the square of the 
trace of $H$
\begin{align}
{\rm tr}(H^2) = \sum_{jk} \la H_{jk} H_{kj}\ra = \frac{N^2}{N} = N, \\
\big [\, {\rm tr}(H)\, \big ]^2 = \sum_{jk} \la H_{jj} H_{kk}\ra
 = \sum_j \la H_{jj}^2\ra = 1.
\end{align}
In turn, this means that  the eigenvalues $\eps_j$ of $H_\rme$ lie essentially
in the interval $(-2,2)$, and have a  semi-circle distribution, for large 
$N$. In addition, since $\sum_j \la\eps_j^2\ra = N$, 
\begin{align}
\la\eps_j^2\ra = 1, 
\end{align}
and, since $\sum_{jk} \la\eps_j\eps_k\ra = N + N(N-1) \la\eps_j\eps_k\ra = 1$,
\begin{align}
\la\eps_j\eps_k\ra_{j\ne k} = -\frac{1}{N}.
\end{align}
\section{The Gaussian PUE} 
\label{cap:PUE:promedios}
Initial calculations were done in a Poissonian ensemble with Gaussian level
density. This could correspond to spin models. Even though non-Markovian
effects are not visible here (the Fourier transform of a Gaussian, is another
Gaussian), some results are easier to obtain, and provide a guide to the more
complicated calculations in the GUE case.  We present some details here, which
might also provide a guide for the calculations using other spectral densities. 

We now assume that the coupling $V$ is taken from the GPUE (Gaussian PUE). This
means that $V$ is chosen as
\begin{equation}
V_{\rm GPUE} =U D U^{\dagger}\,,
\end{equation}
with  $U$ chosen from the unitary group with the Haar measure, and $D$ is a
diagonal matrix with Gaussian independent numbers with $\sigma =1$. 
This means that the variables $v_{\mu}$ in \Eref{eq:A1} are, again, independent
Gaussian variables with mean zero and unit standard deviation. 

The subsequent calculation runs in identical way as shown in \Sref{sec:strong}, 
except that in \Eref{cap:GUE:eq:charact_sc}, we have 
\begin{equation}
b_1(t) = \rme^{-\frac{t^2}{2}}, 
\label{}
\end{equation}
and one has to propagate this difference through \Eref{cap:GUE:eq:aux}. 
In this case, some of the integrals involved in the terms 
$F^{(i)}_\nu$ can be performed, though not all.

In this case, the compact expression for the large dimension limit, 
	\begin{equation}
	\alpha_2^\infty (t) \approx \sqrt{\pi} t \rme^{-\frac{3 t ^2}{4}} \mathrm{Erf}\left(\frac{t}{2}\right)\,-\,\pi \rme^{-\frac{t^2}{2}}\mathrm{Erf}\left(\frac{t}{2}\right)^2,
	\end{equation}
is obtained. 

\bibliographystyle{apsrev} 
\bibliography{refs}
\end{document}